\documentstyle[12pt]{article}
\begin {document}

\title {ON THE EFFECTIVE INERTIAL MASS DENSITY OF A DISSIPATIVE FLUID}

\author{L. Herrera\thanks{Postal address: Apartado 80793, Caracas 1080A, Venezuela} 
\thanks{e-mail address: laherrera@telcel.net.ve}\\Escuela de F\'\i
sica. Facultad de Ciencias.\\ Universidad Central de Venezuela, Caracas, Venezuela}   

\date{}
\maketitle

\begin{abstract}
It is shown that the effective inertial mass density of a dissipative
fluid just after leaving the equilibrium, on a time scale of the order of
relaxation time, reduces by a factor which depends on dissipative
variables. Prospective applications of this result to cosmological and
astrophysical scenarios are discussed. 
\end{abstract}

\newpage

\section{Introduction}
In classical dynamics the inertial mass is defined as the factor of
proportionality between the three-force applied to a particle (a fluid
element) and the resulting three-acceleration, according to Newton's
second law.

In relativistic dynamics a similar relation only holds (in general) in the
instantaneous rest frame (i.r.f.), since the three-acceleration and the
force that causes it are not (in general) paralell, except in the i.r.f.
(see for example \cite{Ri}).

In this work we shall derive an expression for the effective inertial
mass density of a dissipative relativistic fluid, which is valid just
after the fluid leaves the equilibrium, on a time scale of the order of
relaxation time. 

By ``effective inertial mass'' (e.i.m.) density we mean the factor of
proportionality between the applied three-force density and the resulting
proper acceleration (i.e., the three-acceleration measured in the i.r.f.).

As we shall see, the obtained expression for the e.i.m. density contains
a contribution from dissipative variables which reduces its value with
respect to the non-dissipative situation. Such decreasing of e.i.m.
density has already been brought out in the spherically symmetric
self-gravitating case \cite{Hetal}, the axially symmetric
self-gravitating case \cite{HeDP} and also for slowly rotating
self-gravitating systems \cite{HM98}. Here we want to provide a general
derivation for the e.i.m. density of a dissipative fluid, which is independent on the symmetry of the problem.On the other hand,
in order to derive the expression for the e.i.m. density, it will be necessary to suppress gravitational contributions to four acceleration. This will be achieved
by evaluating the equation of motion in a locally Minkowski frame, where local gravitational effects vanish. Then evaluating
the resulting equation of motion in the i.r.f. we are led to an expression which takes the desired, "Newtonian", form.
$$
\mbox{Force=e.i.m.}\times\mbox{acceleration(proper),}
$$
 It is perhaps worth noticing that
the concept of effective inertial mass is familiar in other branches of physics, thus
for example the e.i.m. of an electron moving under a given force through
a crystal, differs from the value corresponding to an electron moving
under the same force in free space, and may even become negative (see
\cite{Ja,Ki}).

In order to illustrate our point and to establish the notation, we shall 
first consider, very briefly, the perfect fluid case. However before closing this Section, the following comments are in order:

It is worth noticing, that there is not a unique way to define the four velocity of a dissipative fluid. Indeed, we may take,  among other possibilities,  the
four velocity vector as being related  to the velocity of energy transport (Landau-Lifshitz \cite{LL}) or  to the velocity of particles (fluid element) as done by
Eckart (\cite{Eckart}), both choices are of course perfectly equivalent, and both lead to different definitions of rest frame. In the former case there is not
energy flux in the rest frame (the $T^{0j}$ components of the energy tensor vanish), whereas in the later the spatial components of the particle current four
vector vanish in the comoving frame. In this work we adopt the Eckart choice.

This is motivated by the fact that  the very concept of e.i.m. as defined above, requires (and makes sense
only for)  such  choice (Eckart).  Indeed, observe that the concept of e.i.m. requires  the existence of a frame where the three-force applied to a particle and the
resulting three acceleration are parallel, and this only happens in the i.r.f. where a given fluid element is at rest. But this is the
Eckart frame by definition.

In the Landau-Lifshitz frame we cannot say that such relation between the applied three-force and the resulting three accelerations follows, and therefore in such
case the very meaning of e.i.m. is uncertain.

\section{The e.i.m. density of a relativistic perfect fluid}
Let us consider a relativistic perfect fluid, whose energy-momentum
tensor takes the usual form
\begin{equation}
T^{\mu\nu}=(\rho + p) u^\mu u^\nu - p g^{\mu\nu}
\label{Tpf}
\end{equation}
(we use relativistic units and signature $-2$)

Then the equation of motion yields
\begin{equation}
(\rho + p) a^\alpha = h^{\alpha\nu} p_{,\nu}
\label{em}
\end{equation}
with
\begin{equation}
h^\alpha_\mu \equiv \delta^\alpha_\mu - u^\alpha u_\mu
\label{h}
\end{equation}
\begin{equation}
\dot u^\mu \equiv a^\mu = u^\nu u^\mu_{;\nu}
\label{a}
\end{equation}
where colon and semicolon denote partial and covariant derivatives,
respectively. 

Observe that for the sake of generality we are including gravitational
field, although, as already noticed, the final expression for e.i.m. density will be
independent of gravity.

Eq.(\ref{em}) suggests that $(\rho + p)$ plays the role of the e.i.m.
density of the fluid (which is in fact the case). However, we recall that
the applied three-force and the resulting three-acceleration are paralell
only in the i.r.f., therefore, in order to conclude that $(\rho + p)$
represents the e.i.m. density, (\ref{em}) has to be evaluated in such a
frame. Furthemore, it should be observed that $a^\alpha$ contains both,
gravitational (in the self-gravitating case) and ``kinematical''
contributions. Since in the concept of e.i.m. only the later should
appear, we have to evaluate (\ref{em}) not only in the i.r.f., but also
in the locally Minkowskian frame (l.m.f.) where local gravitational effects
vanish. Formally, this may be achieved, either by introducing locally
Minkowskian coordinates or equivalently, by considering a tetrad field
attached to such l.m.f.. Denoting by a tilde the components of tensors in the l.m.f.,equation (\ref{em}) reads 
\begin{equation}
(\rho + p) \tilde{a}^{\mu} = \tilde{h}^{\mu\nu} p_{,\nu}
\label{tc2}
\end{equation}

Evaluating (\ref{tc2}) in the i.r.f., it becomes clear that $(\rho + p)$
defines the e.i.m. density of any fluid element (a well known fact, see
for example \cite{Li}). 

In order to better illustrate this point, let us
consider the spherically symmetric case.

Thus, let the line element be
\begin{equation}
ds^2 = e^\nu dt^2 - e^\lambda dr^2 - r^2 (d\theta^2 + \sin^2{\theta}
d\phi^2)
\label{le}
\end{equation}
with $\nu = \nu(r,t)$ and $\lambda = \lambda(r,t)$, then locally Minkowski coordinates may be introduced by \cite{Bondi}
\begin{equation}
dT=e^{\nu/2}dt\,\qquad\,dx=e^{\lambda/2}dr\,\qquad\,
dy=rd\theta\,\qquad\, dz=rsin\theta d\phi
\label{Min}
\end{equation}
satisfying
\begin{equation}
ds^2 = dT^2 - dx^2 - dy^2 - dz^2
\label{ellin}
\end{equation}
Next, the four-velocity vector for any  fluid element, in the frame of
(\ref{le}) is 
\begin{equation}
u^\mu = \left(\frac{e^{-\nu/2}}{\sqrt{1-\omega^2}},
\frac{\omega e^{-\lambda/2}}{\sqrt{1-\omega^2}}, 0, 0\right)
\label{um}
\end{equation}
where $\omega$ is the radial velocity of the fluid element as measured in
the l.m.f.

Therefore, in the l.m.f. the components of  (\ref{um}) are
\begin{equation}
\tilde{u}^{\mu} = 
\frac{\delta^{\mu}_{T}}{\sqrt{1-\omega^2}} +
\frac{\omega \delta^{\mu}_{x}}{\sqrt{1-\omega^2}} 
\label{tum}
\end{equation}
and the components of the four-acceleration in the l.m.f. read
\begin{equation}
\tilde{a}_{\mu} = \tilde{u}^{\nu} \tilde{u}_{\mu,\nu}
\label{ta}
\end{equation}
Then, feeding back (\ref{ta}) into (\ref{tc2}), and evaluating in the
i.r.f. ($\omega=0$), we obtain the ``Newtonian-type'' equation
\begin{equation}
(\rho + p) \omega_{,T}= - p_{,x} 
\label{nt}
\end{equation}
indicating that $(\rho + p)$ represents the e.i.m. density of the fluid.

Let us now turn to the dissipative case.

\section{The e.i.m. density of a dissipative fluid}
For simplicity we shall consider the case of pure heat conduction,
neglecting viscosity terms. Then, the
energy-momentum tensor takes the usual form
\begin{equation}
T_{\mu\nu} = (\rho + p) u_\mu u_\nu - p g_{\mu\nu} + q_\mu u_\nu + q_\nu
u_\mu
\label{Tdf}
\end{equation}
and the equation of motion, projected onto the plane orthogonal to $u^\mu$
reads
\begin{equation}
(\rho + p) a^\alpha - p_{,\nu} h^{\nu\alpha} + h^\alpha_\mu q^\mu_{;\nu}
u^\nu + q^\alpha u^\nu_{;\nu} + q^\nu u^\mu_{;\nu} h^\alpha_\mu = 0
\label{emd}
\end{equation}
where $q^\mu$, satisfying $q^\mu u_\mu = 0$, denotes the heat flow vector.

We have now to adopt a transport equation for $q^\mu$. Although this is
still a matter of discussion, it should be borne in mind that some kind
of hyperbolic dissipative theory \cite{JCL} should be applied in order to
avoid violation of causality and other undesirable consequences derived
from the (parabolic) Eckart-Landau approach. Furthermore, since we are
interested in the transient regime between two equilibrium states we are
compelled to use a hyperbolic theory, for it is known that the parabolic
theory assumes that the system is relaxed at all times.

Thus, we shall consider the Israel-Stewart transport equation \cite{JCL}
(although any hyperbolic equation yielding a Cattaneo-type \cite{JCL}
equation in the non-relativistic limit would lead to the same result)
\begin{equation}
\tau h^\mu_\nu \dot q^\nu + q^\mu = \kappa h^{\mu\nu} (T_{,\nu} - T
a_\nu) - \frac{1}{2} \kappa T^2 \left(\frac{\tau u^\alpha}{\kappa
T^2}\right)_{;\alpha} q^\mu + \tau \omega^{\mu\nu} q_\nu
\label{tr}
\end{equation}
where $\tau, \kappa, T$ and $\omega^{\mu\nu}$ denote the relaxation time,
the thermal conductivity, the temperature and the vorticity tensor,
respectively.

Then, feeding back (\ref{tr}) into (\ref{emd}) one obtains
\begin{eqnarray}
&& a^\alpha \left(\rho + p - \frac{\kappa T}{\tau}\right) - p_{,\nu}
h^{\nu\alpha} - \frac{1}{\tau} \left(q^\alpha - \kappa
h^{\alpha\nu} T_{,\nu}\right) \nonumber \\
& - & \frac{1}{2 \tau} \kappa T^2 
\left(\frac{\tau u^\beta}{\kappa
T^2}\right)_{;\beta} q^\alpha 
 +  2 \omega^{\alpha\nu} q_\nu +
q^\alpha
\Theta + q^\nu \sigma^\alpha_\nu = 0 
\label{dis}
\end{eqnarray}
with
$$
\Theta = u^\nu_{;\nu}
$$
$\sigma_{\mu\nu}$ as usual denotes the shear tensor, and where the
relation \cite{De}
\begin{equation}
u_{\alpha;\beta} = \omega_{\alpha\beta} + \sigma_{\alpha\beta} -
\frac{1}{3} \Theta h_{\alpha\beta} - a_{\alpha} u_{\beta}
\label{rel}
\end{equation}
has been used.

We shall now consider that our system is initially, either in hydrostatic
equilibrium without dissipation, or very close to equilibrium with
``small'' dissipation (quasi-stationary regime). Next, let us suppose
that at some moment (say $t = \tilde t$) it leaves the initial state
entering into a dissipative regime.

Then, ``immediately'' after departure from equilibrium (or
quasi-equilibrium), where ``immediately'' means on a time scale of the
order of $\tau$ (or smaller), the last four terms in (\ref{dis}) can be
neglected, and we have 
\begin{equation}
a^\alpha \left(\rho + p - \frac{\kappa T}{\tau}\right) - p_{,\nu}
h^{\nu\alpha} - \frac{1}{\tau} \left(q^\alpha - \kappa
h^{\alpha\nu} T_{,\nu}\right)  = 0 
\label{af}
\end{equation}

The components of (\ref{af}) in the l.m.f. are
\begin{equation}
\tilde{a}^{\mu} \left(\rho + p - \frac{\kappa T}{\tau}\right) - p_{,\nu}
\tilde{h}^{\mu\nu} - \frac{1}{\tau} \left(\tilde{q}^{\mu} - \kappa
\tilde{h}^{\mu\nu} T_{,\nu}\right)  = 0
\label{afte}
\end{equation}

However, in the l.m.f., the  components of the heat flux vector in the i.r.f., in
the quasi-stationary regime, satisfy the ``Maxwell-Fourier'' equation
\begin{equation}
\tilde{q}^{\mu} = \kappa \tilde{h}^{\mu\nu} T_{,\nu}
\label{mfq}
\end{equation}

Indeed, just after leaving the equilibrium, the absolute value of the heat flow vector is the same as that corresponding to the quasistationary regime, and
furthermore in that regime (in the l.m.f. and in the i.r.f.) both the kinematical and gravitational contributions to the four acceleration vanish, thereby
justifying (\ref{mfq}).Therefore , just after leaving the quasi-stationary regime, the last term in  (\ref{afte}) vanishes, and this equation becomes
\begin{equation}
\tilde{a}^{\mu} \left(\rho + p - \frac{\kappa T}{\tau}\right) = p_{,\nu}
\tilde{h}^{\mu\nu} 
\label{afmf}
\end{equation}
implying that the e.i.m. density is given by
\begin{equation}
\rho + p - \frac{\kappa T}{\tau}
\label{eimd}
\end{equation}
or,
\begin{equation}
(\rho+p) ({1-\alpha})
\label{nueva}
\end{equation}
with
\begin{equation}
\alpha \equiv \frac{\kappa T}{\tau (\rho + p)}
\label{alpha}
\end{equation}
\section{Discussion}
We have seen that just after leaving the equilibrium the e.i.m. density
of the dissipative fluid reduces its value (as compared to the
non-dissipative situation) by the factor 
\begin{equation}
{1 - \alpha}
\label{al}
\end{equation}
giving rise, in principle, to the possibility of vanishing ($\alpha=1$)
or even negative ($\alpha>1$) e.i.m. density. As mentioned in the
Introduction, such effect was already found in some specific examples,
here it appears as a general result for any dissipative fluid
(self-gravitating or not).

It should be noticed that  since the system  is evaluated immediately (in the sense explained above) after leaving the equlibrium (or quasi-equilibrium)
then the physical meaning of thermodynamical variables (their  values), are extrapolated from the state of equilibrium, and therefore any discussion about 
difficulties of interpreting such variables in a transient regime, is out of the point.

Also, it should be observed that causality and stability conditions hindering
the system to attain condition $\alpha=1$, are obtained on the basis of a
linear approximation, whose validity, close to the critical point
($\alpha=1$), is questionable \cite{HM97}.

At any rate, examples of fluids attaining the critical point and
exhibiting reasonble physical properties have been presented elsewhere
\cite{HeMaI,HeMaII}.

In order to evaluate $\alpha$, let us turn back to c.g.s. units. Then,
assuming for simplicity $\rho + p \approx 2\rho$, we obtain
\begin{equation}
\frac{\kappa T}{\tau (\rho + p)} \approx \frac{[\kappa][T]}{[\tau][\rho]}
\times 10^{-42}
\label{un}
\end{equation}
where $[\kappa]$, $[T]$, $[\tau]$, $[\rho]$ denote the numerical values of
these quantities in $erg. \, s^{-1} \, cm^{-1} \, K^{-1}$, $K$, $s$ and
$g.\, cm^{-3}$, respectively.

Obviously, this will be a very small quantity (compared to $1$), unless
conditions for extremely high values of $\kappa$ and $T$ are attained. In
this respect, observe that although small values of $\tau$ increases
(\ref{un}), they are of little help since (\ref{eimd}) is valid only on a
time scale of the order of $\tau$. Therefore, a decreasing of e.i.m.
density with physically relevant consequences requires values of
(\ref{un}) close to the unit, due to large values of $\kappa$ and $T$ but
non-negligible values of $\tau$.

At present we may speculate that  $\alpha$ may
decrease substantially (for non-negligible values of $\tau$) in a pre-supernovae event

Indeed, at the last stages of massive star evolution, the decreasing of the opacity
of the fluid, from very high values preventing the propagation of photons
and neutrinos (trapping \cite{Ar}), to smaller values, gives rise to
radiative heat conduction. Under these conditions both $\kappa$ and $T$
could be sufficiently large as to imply a substantial increase of
$\alpha$. Indeed, the values suggested in \cite{Ma} ($[\kappa] \approx
10^{37}$;
$[T] \approx 10^{13}$; $[\tau] \approx 10^{-4}$; $[\rho] \approx
10^{12}$ ) lead to $\alpha \approx 1$. The obvious consequence of which
would be to enhance the efficiency of whatever expansion mechanism, of
the central core, at place, because of the decreasing of its e.i.m.
density.
At this point it is worth noticing that the relevance of relaxational effects on gravitational collapse has been recently exhibited and stressed (see
\cite{Collapse}, and references therein)

It is also worth noticing that  the inflationary equation of state
(in the perfect fluid case) $\rho + p = 0$, is, as far as the equation of
motion is concerned, equivalent to $\alpha = 1$ in the dissipative case (both imply the vanishing of the e.i.m. density).

Also observe,  that if we impose the equation of state  $\rho + p = 0$, in the dissipative case,
then the  resulting e.i.m. density is negative (-$\frac{\kappa T}{\tau}$), implying that the attractive gravitational force leads to an overall expansion.

Finally, it is worth stressing that it is the first term on the left in (\ref{tr}), the direct responsible for the decreasing in the e.i.m density, all terms
including kinematical variables vanishing within the time scale considered (after leaving equilibrium). Therefore any hyperbolic dissipative theory yielding a
Cattaneo-type equation in the non-relativistic limit, is expected to give a result similar to the one obtained here.

\section*{Acknowledgements}

The financial assistance from  M.C.T.,Spain, BFM2000-1322 is
gratefully acknowledged.

\end{document}